\documentclass[%
aip,
cp,  
amsmath,amssymb,
reprint,%
]{revtex4-2}

\usepackage{graphicx}
\graphicspath{{./figures/}}
\usepackage{float}
\usepackage{subfig}
\usepackage[labelfont=bf]{caption}
\usepackage{dcolumn}
\usepackage{bm}

\usepackage[utf8]{inputenc}
\usepackage[T1]{fontenc}
\usepackage{mathptmx} 

\begin{document}
	
	\title{Mixed Convection in a Differentially Heated Cavity
		with Local Flow Modulation via Rotating Flat Plates}
	
	\author{Md. Azizul Hakim} 
	\affiliation{Department of Mechanical Engineering, Bangladesh University of
		Engineering and Technology, Dhaka-1000, Bangladesh}
	
	\author{Atiqul Islam Ahad}%
	\affiliation{Department of Mechanical Engineering, Bangladesh University of
		Engineering and Technology, Dhaka-1000, Bangladesh}
	
	\author{Abrar Ul Karim}
	\affiliation{Department of Mechanical Engineering, Bangladesh University of
		Engineering and Technology, Dhaka-1000, Bangladesh}
	
	\author{Mohammad Nasim Hasan}
	\email[Corresponding author: ]{nasim@me.buet.ac.bd}
	\affiliation{Department of Mechanical Engineering, Bangladesh University of
		Engineering and Technology, Dhaka-1000, Bangladesh}%
	\date{\today} 
	
	\begin{abstract}
		Mixed Convection inside a cavity resulting from thermal buoyancy force under
		local modulation via rotating flat plate has been investigated. The present model
		consists of a square cavity with the left and right vertical walls fixed at
		constant high and low temperatures respectively while the top and bottom walls
		are supposed to be adiabatic. Two clockwise rotating flat plates, having negligible thickness in
		comparison to their lengths, acting as flow modulators have been placed
		vertically along the centerline of the cavity. The moving boundary problem due to
		plate motion in this study has been solved by implementing \textit{Arbitrary
			Lagrangian Eulerian (ALE)} finite element formulation with triangular
		discretization scheme.Simulations are conducted for air ($Pr=0.71$) at different
		Rayleigh numbers ($10^2 \leq Ra \leq 10^6$). Rotational Reynolds number based on
		plate dynamic condition has been considered to be constant at 430. Numerical
		results identify critical Rayleigh number $Ra_{cr}=0.41\times10^6$ beyond which
		two smaller flow modulators are more effective than a single larger modulator.
		Thermal oscillating frequency was observed to be insensitive to Rayleigh number
		for the case of double modulators.
	\end{abstract}
	
	\maketitle
	
	\section{INTRODUCTION}
	
	Investigations of mixed convection heat transfer have a great number of empirical
	applications in lubrication technologies, oil pipelines, electronic compound
	cooling, etc. These studies have significant importance in the designs of
	electronic devices to dissipate heat from them to avoid thermal damages. Fu
	\textit{et al.} [2], Kimura \textit{et al.} [4] and Ghad-dar and Thiele [3] took
	the primary steps to investigate the heat transfer in a square cavity owing to
	thermal buoyancy force under the effect of rotating cylinder. Lewis [1] numerically
	investigated steady flow in a square cavity with a rotating cylinder for $1 \leq
	Re \leq 1400$. Khanafer and Aithal [5] numerically studied mixed convection heat
	transfer in a lid-driven cavity having a rotating cylinder for various system parameters
	for example: Richardson number and dynamic condition of the cylinder. Their findings revealed that both the thermal 
	and flow field inside the cavity depended heavily on the speed and direction of rotation of cylinder.
	Billah \textit{et al.} [6] investigated mixed convection heat transfer in a
	channel with active flow modulation via rotating cylinder. They found that type of configuration and
	direction of cylinder rotation strongly influenced the heat transfer. The heat transfer performance as represented
	by average Nusselt number over the heat source had been found to show an increasing
	trend with Reynolds and Grashoff numbers but at higher values of Reynolds and Grashoff numbers, the influence of local flow
	modulation became weak. The experimental work of Kimura \textit{et
		al.} [7] used a rotating blade in a square enclosure as a heat transfer
	augmentor. It had been realized that the rotating blade is more effective than
	rotating cylinder in strengthening the heat transfer. Lee \textit{et al.} [8] led
	similar investigation numerically and reported that thermal oscillation appeared
	when Rayleigh number went beyond a critical value.
	
	As delineated in the literature survey outlined above, a numerical study of laminar mixed
	convection heat transfer in a differentially heated cavity with single flow
	modulator had been carried out [8] but effects of multiple smaller flow modulators in
	a differential heated cavity has not been investigated yet. The present study
	aims to explore the effect of Rayleigh ($Ra$) number on the flow pattern and heat
	transfer in a differentially heated cavity with multiple flow modulators.Results obtained in this study are discussed in terms of streamlines, isothermal contours, heatlines, spatially averaged Nusselt number and 
	time averaged Nusselt number at the heated wall of the cavity.
	
	\section{PROBLEM DESCRIPTION}
	The configuration under study, shown in Fig. 1, is a square cavity of dimension
	$L$ with two rotors along the centerline of the cavity having lengths $d(=0.3L)$
	and negligible thickness. The left wall is maintained at a constant high
	temperature (\textit{$T_h$}) and the right wall is kept at low temperature
	(\textit{$T_c$}) while the upper and lower walls are assumed to be insulated. Thermal resistance
	across the modulators is negligible as the thickness of the rigid rotors are
	assumed to be negligible. The cavity is completely filled with air and assumed
	to be a Newtonian fluid. The modulators are assumed to rotate at a constant frequency in
	clockwise direction to match the sense of natural convection.
	\begin{figure}[H]
		\begin{center}
			\includegraphics[width=3in, height=3in]{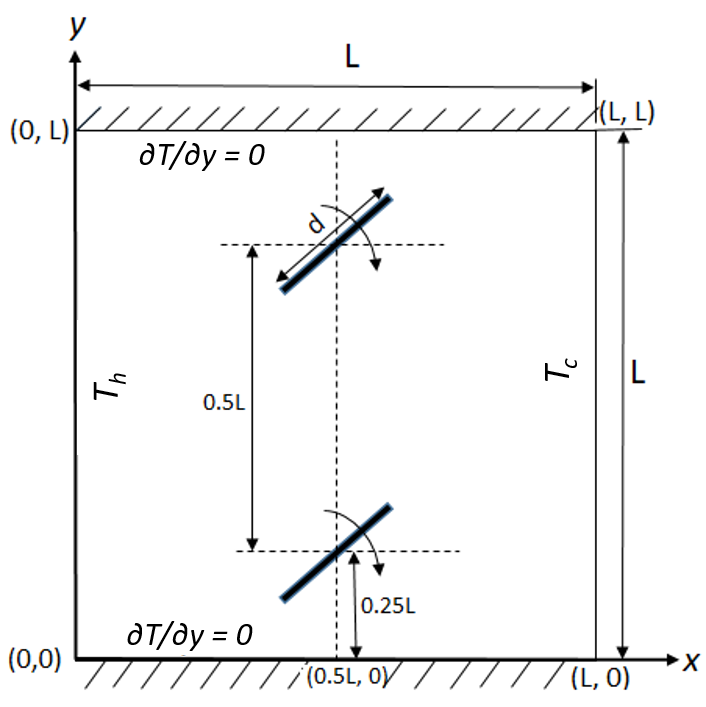}
			\caption{Schematic of the present problem.}
		\end{center}
	\end{figure}
	Within the cavity, the flow has been assumed to be laminar, incompressible and
	two-dimensional. The Boussinesq approximation is used to define the thermal
	buoyancy force. All other thermophysical fluid properties are assumed to be
	constant. In the energy equation, the viscous dissipation term is neglected.
	Taking into consideration the assumptions stated above, the non-dimensional
	conservation equations of momentum, mass and energy for mixed convention in 2-D
	Cartesian coordinate system can be expressed as:
	\begin{equation}
	\frac{\partial U}{\partial X}+\frac{\partial V}{\partial Y}=0
	\end{equation}
	\begin{equation}
	\frac{\partial U}{\partial \tau}+U\frac{\partial U}{\partial X}+V
	\frac{\partial U}{\partial Y}=-\frac{\partial P}{\partial X}+\frac{1}{Re}
	\left(\frac{\partial^2 U}{\partial X^2}+\frac{\partial^2 U}{\partial Y^2}\right)
	\end{equation}
	\begin{equation}
	\frac{\partial V}{\partial \tau}+U\frac{\partial V}{\partial X}+V
	\frac{\partial V}{\partial Y}=-\frac{\partial P}{\partial Y}+\frac{1}{Re}
	\left(\frac{\partial^2 V}{\partial X^2}+\frac{\partial^2 V}{\partial Y^2}\right)+
	\frac{Ra}{Re^2Pr}\theta
	\end{equation}
	\begin{equation}
	\frac{\partial \theta}{\partial \tau}+U\frac{\partial \theta}{\partial X}+V
	\frac{\partial \theta}{\partial Y}=\frac{1}{RePr}\left(\frac{\partial^2 \theta}
	{\partial X^2}+\frac{\partial^2 \theta}{\partial Y^2}\right)
	\end{equation}
	\\
	Equations (1)-(4) have been normalized using the following dimensionless scales:
	$$X=\frac{x}{L}, Y=\frac{y}{L}, U=\frac{u}{u_c}, V=\frac{v}{u_c}, P=\frac{p}{p_{ref}}, \theta =\frac{T-T_c}{T_h-T_c}, \tau=\frac{t}{f^{-1}} $$\\
	Here, $x$ is the distance along horizontal direction and $y$ is that along the
	vertical direction; $u$ and $v$ are the horizontal and vertical components of
	velocity respectively; $p$ is pressure, $T$ is temperature. The reference
	quantities are length $L$, velocity $u_c=fL$ where $f$ is the rotational
	frequency,reference pressure $p_{ref}=\rho u_{c}^2$ and temperature difference $\Delta T = T_h – T_c 
	$. $\alpha$, $\beta$, $\rho$, $v$ are thermal diffusivity, coefficient of
	volumetric expansion, fluid density and kinematic viscosity respectively. In the
	above equations, the governing parameters Prandtl number ($Pr$), Grashof number
	($Gr$), rotational Reynolds number ($Re$) and Rayleigh number ($Ra$) have been
	defined as follows:
	$$Gr=\frac{g\beta (T_h-T_c)H^3}{v^2}, Pr=\frac{v}{\alpha}, Re=\frac{u_c H}
	{v}, Ra=GrPr$$
	The associated boundary conditions are as follows:
	$$U(0,Y)=0, V(0,Y)=0, \theta (0,Y)=1$$
	$$U(1,Y)=0, V(1,Y)=0, \theta (1,Y)=0$$
	$$U(X,0)=0, V(X,0)=0, \partial \theta (X,0)/\partial Y=0$$
	$$U(X,1)=0, V(X,1)=0, \partial \theta (X,1)/\partial Y=0$$
	No-slip condition has also been implied on the blade surfaces. Heat flow can be
	visualized better in terms of the heat function ($\Pi$) obtained from conductive
	heat fluxes as well as from convective fluxes ($U\theta,V\theta$) as defined
	below:
	\begin{equation}
	\frac{\partial \Pi}{\partial Y}=U\theta-\frac{1}{RePr}\frac{\partial \theta}
	{\partial X}
	\end{equation}
	\begin{equation}
	-\frac{\partial \Pi}{\partial X}=V\theta-\frac{1}{RePr}\frac{\partial \theta}
	{\partial Y}
	\end{equation}
	Local Nusselt number is defined as:
	\begin{equation}
	Nu(X,\tau)=-\left(\frac{\partial \theta}{\partial X}\right)_{X=0}
	\end{equation}
	Spatial averaged Nusselt number is obtained after integrating the local Nusselt
	number along the left hot wall of the cavity:
	\begin{equation}
	Nu(\tau)=\frac{1}{L}\int_{0}^{1} Nu(Y,\tau)dY
	\end{equation}
	Both time and spatial averaged Nusselt number is calculated for one period of
	oscillation:
	\begin{equation}
	Nu_{avg}=\int_{0}^{1/ft} Nu(\tau)d\tau
	\end{equation}
	For heat function ($\Pi$), at adiabatic top and bottom walls, Dirichlet boundary
	condition $\Pi=0$ applies and for isothermal sidewalls a Neumann boundary
	condition ($n.\Delta \Pi=0$) applies.
	
	\subsection{NUMERICAL METHOD}
	The numerical method that has been implemented to solve the governing equations
	(1)-(4) of the present moving mesh problem is based on \textit{Arbitrary Lagrangian Eulerian (ALE)} finite element formulation with non-uniform triangular
	discretization scheme. The non-dimensional partial differential equations of
	fluid field (1)-(4) along with respective boundary conditions are dicretized by
	Galerkin finite element method. In order to achieve better convergence of the
	calculations in minimum time, a variable time step method has been implemented.
	Moving boundary has been solved on fixed non-staggered Cartesian grids.
	Glowinski [9] and Hu [10] have studied various applications of this procedure in
	solving fluid-solid systems of moving rigid bodies. We used non-linear parametric
	solution technique to solve the governing equations because it helps to converge
	rapidly. Several numerical runs at high $Ra=10^6$ and $Pr=0.71$ for double
	modulator with $d=0.3L$ have been performed to obtain optimum grid distribution
	with accurate results and minimal computational time. Table. 1 summarizes the
	comparison of average Nusselt numbers with various grid sizes. Approximately
	43538 elements ensure that the solution can be taken as mesh independent. In our
	study, at regions near the walls and the blade surfaces, mesh has been refined
	to investigate the swift changes in dependent variables.
	The present numerical procedure of moving mesh has been ascertained against the
	existing results of Lee \textit{et al.}[8]. Fig.2 represents the comparison
	between the variation of Nusselt number with dimensionless time for single
	modulator $d = 0.6L$ and $Re = 430$ for four representative Rayleigh numbers ($Ra
	$).
	\begin{figure}[H]
		\begin{center}
			\includegraphics[width=3in, height=3in]{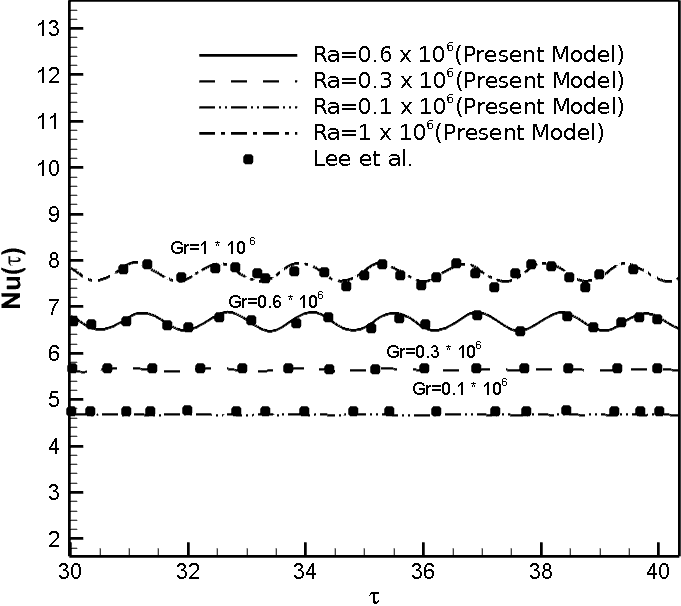}
			\caption{Validation of the present model against Lee \textit{et al.}  [8] in terms of transient average Nusselt number variation.}
		\end{center}
	\end{figure}
	
	\begin{table}[H]
		\caption{Grid independency check in terms of time averaged Nusselt number at Left Wall.}
		\begin{ruledtabular}
			\begin{tabular}{ccccccc}
				Elements & 22618 & 35158 & 40104 & \textbf{43538} & 47824
				& 56082\\\hline
				$Nu_{avg}$ & 7.88 & 8.28 & 8.17 & \textbf{8.26} & 8.25
				& 8.30\\
			\end{tabular}
		\end{ruledtabular}
	\end{table}
	
	\section{RESULTS AND DISCUSSION}
	Present numerical computation has been initiated from a fictional initial
	condition. When periodic steady state is reached, the time is set as $\tau=0$ and
	computed 10 more cycles ($0<\tau<10$). The prime attention has been paid to the
	effects of the number of modulators present and Rayleigh number ($Ra$) on flow
	characteristics and heat transfer within the computational domain. Dynamic
	condition based on rotation of blades ($Re = 430$) is kept fixed and Rayleigh
	number ($Ra$) is varied from $10^2\leq Ra \leq 10^6$. Flow and thermal fields are
	visualized using streamlines, isotherms and heat lines. Fast Fourier Transform
	($FFT$) is computed to find out the thermal oscillating frequency. The heat
	transfer performance within the computational domain is characterized by time and
	spatially averaged Nusselt number ($Nu(\tau)$) along the left heated wall.
	
	\subsection{Heat Transfer Performance Analysis}
	Effect of modulation on the relation between time average Nusselt number
	($Nu_{avg}$) and Rayleigh number ($Ra$) is presented in Fig. 3. It is observed
	that heat transfer with no modulation is greater beyond the point
	$Ra=0.14\times10^6$ for single modulation; below $Ra=4.2\times10^3$ values of
	$Nu_{avg}$ for double modulation is greater than that of no modulation. At $Ra >
	0.41 \times 10^6$, double modulation increases $Nu_{avg}$ more than single
	modulation case due to greater forced convection effect produced by two rotors
	rather than single one. However, these values of $Nu_{avg}$ are lower than that
	of no modulation. This occurs because they can also act as an obstruction in
	mixing of fluid in the zone between the rotors, causing a decrease in natural
	convection. Temporal variation of spatially average Nusselt number for different
	modulation is presented in Fig. 4(a) and (b). From the figure it is clear that
	thermal oscillation is prominent at higher Rayleigh number.
	\begin{figure}[!tbp]
		\begin{center}
			\includegraphics[width=3in, height=3in]{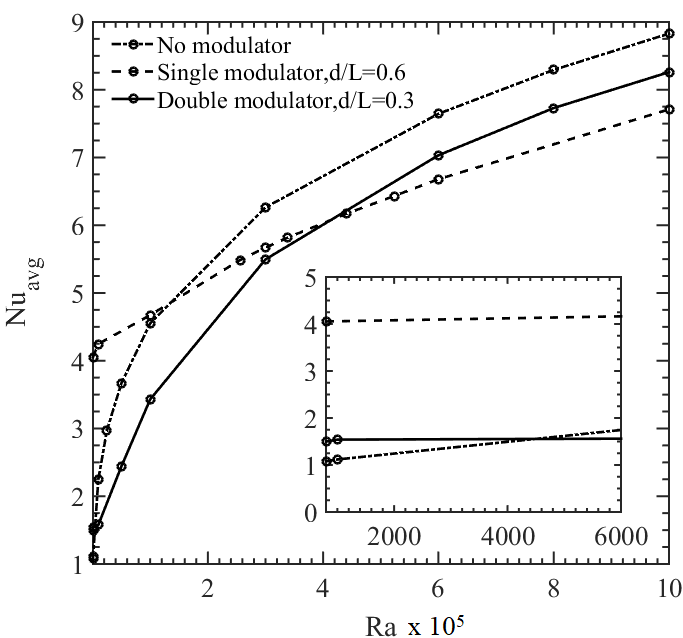}
			\caption{Average Nusselt number($Nu_{avg}$) variation with Rayleigh number for different flow	modulations.}
		\end{center}
	\end{figure}
	\begin{figure}[H]
		\begin{center}
			\subfloat[]{\includegraphics[width=3in, height=3in]{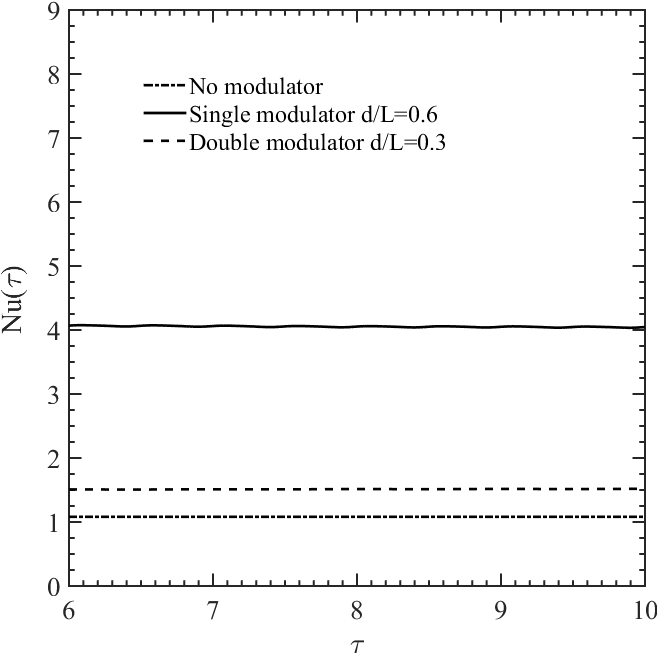}}
			\hspace{0.3cm}
			\subfloat[]{\includegraphics[width=3in, height=3in]{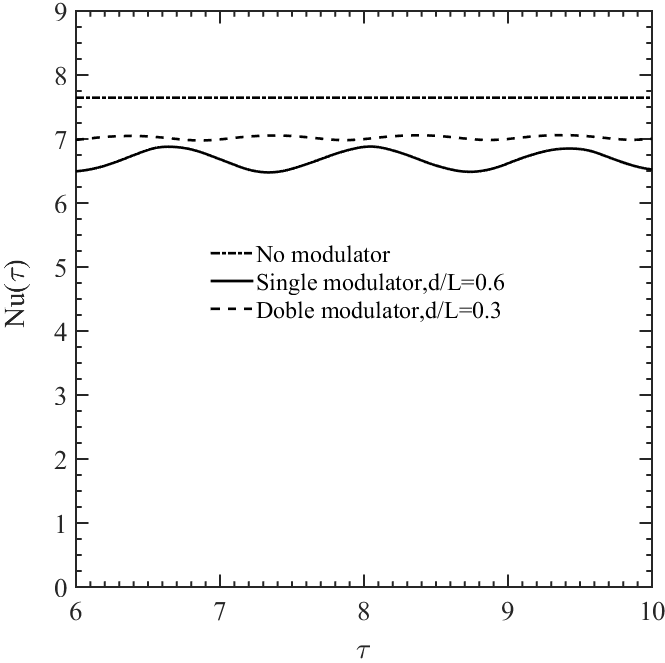}}
			\caption{Temporal Variation of Spatially Averaged Nusselt number for (a)
				$Ra=0.0008\times10^6$ and (b) $Ra=0.6\times10^6$.}
		\end{center}
	\end{figure}
	
	\subsection{Flow Field and Heat Flow Visualization}
	Flow field, isotherms and heatlines for $Ra=0.0008\times10^6(<Ra=4.2\times10^3)$
	and $Ra=0.6\times10^6(>Ra=0.41\times10^6)$ are presented at $\tau =10$ in Fig. 5
	and Fig. 6. This particular dimensionless time is when the flat plates are in
	horizontal position. From the observation of streamlines for cavity without flow
	modulation no vortex is formed. But due to the presence of single modulation four
	vortices can be noticed at the four corners. For two flow modulator several
	vortices can be noticed but they are weaker compared to single modulator. For no
	modulation, parallel and uniform isotherms are noticed
	\begin{figure}
		\begin{center}
			\begin{tabular}{c c c c}
				& No Modulator & Single Modulator, $\frac{d}{L}=0.6$ & Double Modulator, $
				\frac{d}{L}=0.3$\\
				
				\rotatebox{90}{\hphantom{1111111111}Streamlines} & \includegraphics[width=2in,
				height=2in]{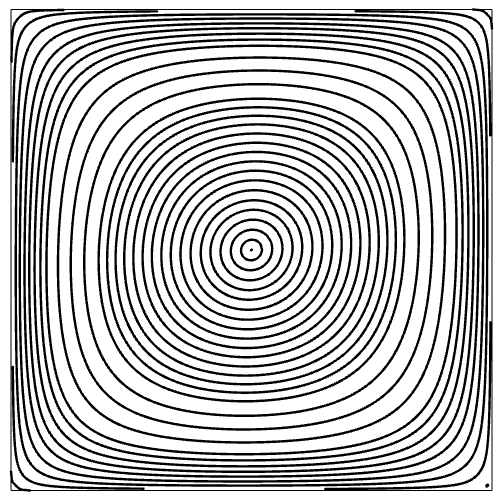} & \includegraphics[width=2in, height=2in]{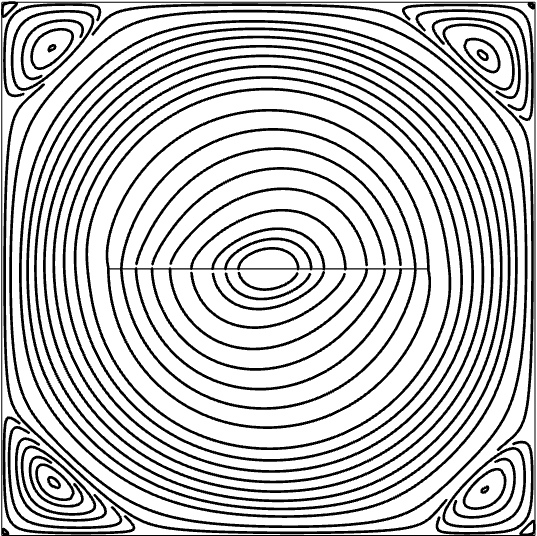} &
				\includegraphics[width=2in, height=2in]{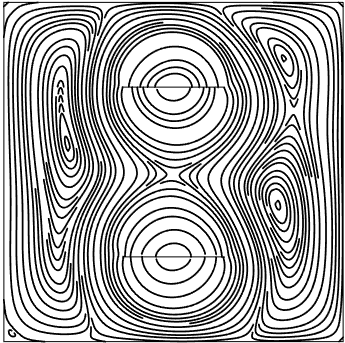}\\
				\rotatebox{90}{\hphantom{1111111111}Isotherms} & \includegraphics[width=2in,
				height=2in]{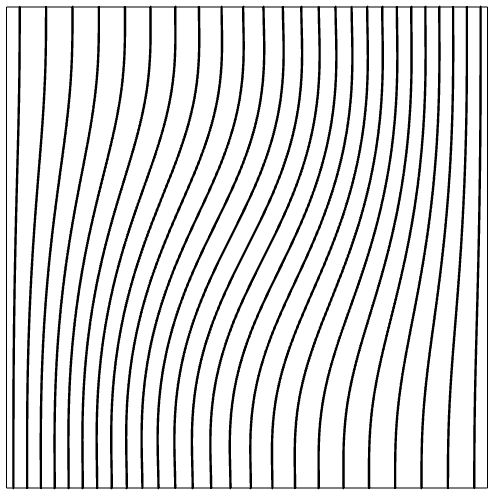} & \includegraphics[width=2in, height=2in]{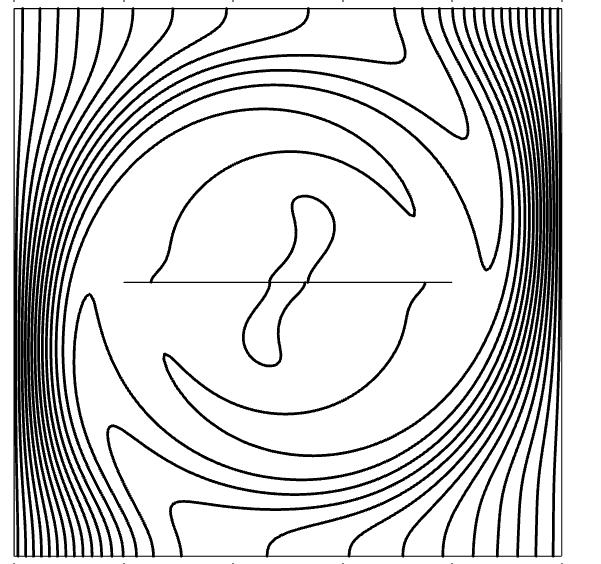} &
				\includegraphics[width=2in, height=2in]{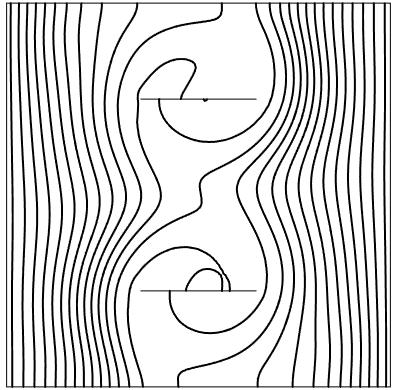}\\
				\rotatebox{90}{\hphantom{1111111111}Heatlines} &
				\includegraphics[width=2in, height=2in]{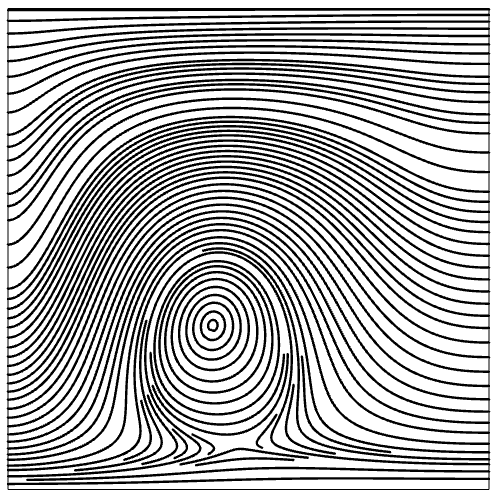} & \includegraphics[width=2in,
				height=2in]{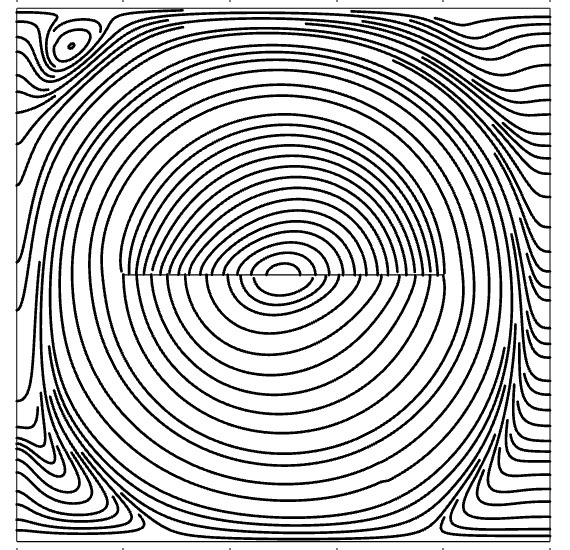} & \includegraphics[width=2in, height=2in]{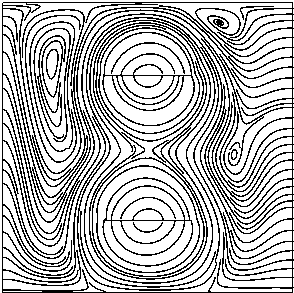}\\
			\end{tabular}
		\end{center}
		\caption{Streamlines, Isotherms and Heatlines for different flow modulations at $Ra=0.0008 \times10^6$ at $\tau=10$}
	\end{figure}
	
	near the side walls. Upon introducing flow modulation, isotherms become more
	clustered towards side walls due to the forced convection induced by rotation in
	the direction of natural convection. Isotherms for double modulator are noticed
	to be aligned parallelly near the vertical walls. From heatlines for no
	modulation, the lines emanating from left hot wall are perpendicular to the
	isotherms, resulting in conduction heat transfer near the side walls. At the
	upper side of the cavity, heat lines are nearly parallel and less dense; hence
	the lesser flow of heat. For single flow modulation the heatlines near the walls
	and inside the cavity tend to bend hence enhancing the energy flow. For $Ra = 0.6
	\times 10^6$ in Fig. 6 streamlines for no modulation involve two vortices. For
	single flow modulation streamlines involve only one vortex at the center. For the
	case of double flow modulation several vortices can be noticed near modulators
	including at their centers. For no modulation, isotherms near the sidewalls are
	observed to be more clustered. Thick thermal boundary layer for no flow
	modulation results in decreasing Nusselt number (about 10.8\% decrease). Two
	modulators tend to push the isotherm toward the left wall hence Nusselt number
	increases (about 6\% increase) compared to single flow modulation.
	\begin{figure}[H]
		\begin{center}
			\begin{tabular}{c c c c}
				& No Modulator & Single Modulator, $\frac{d}{L}=0.6$ & Double Modulator, $
				\frac{d}{L}=0.3$\\
				
				\rotatebox{90}{\hphantom{1111111111}Streamlines} & \includegraphics[width=2in,
				height=2in]{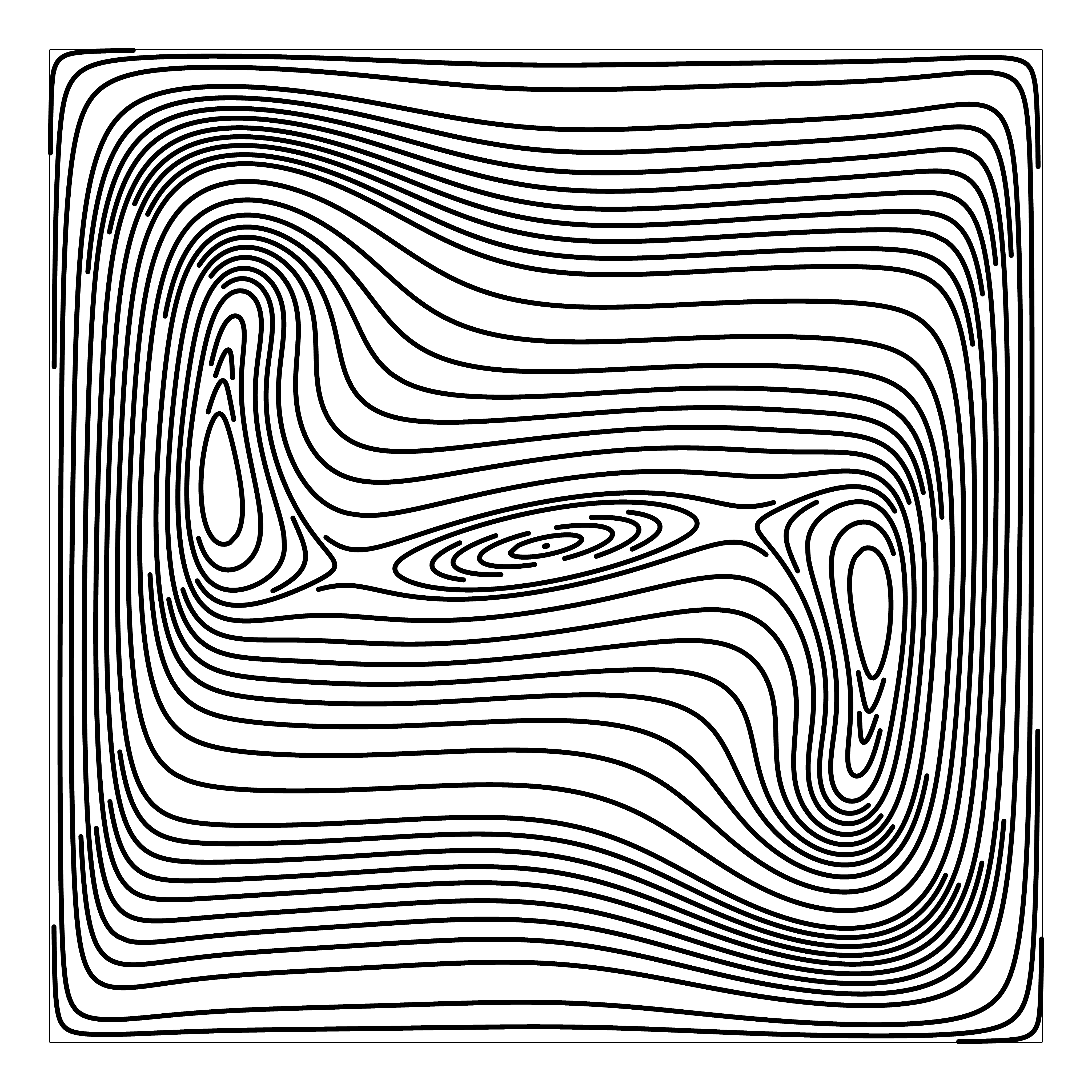} & \includegraphics[width=2in, height=2in]{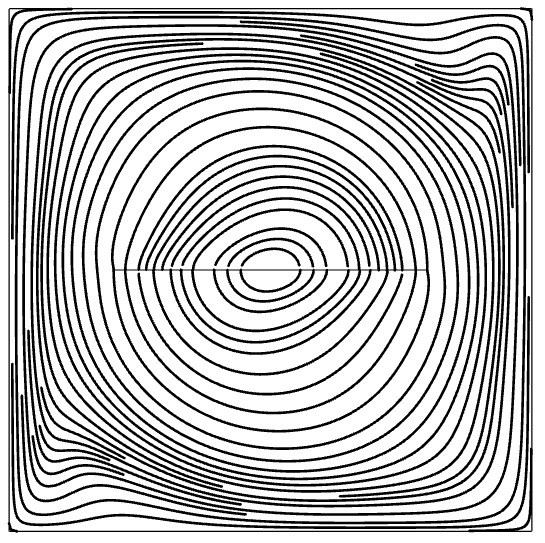} &
				\includegraphics[width=2in, height=2in]{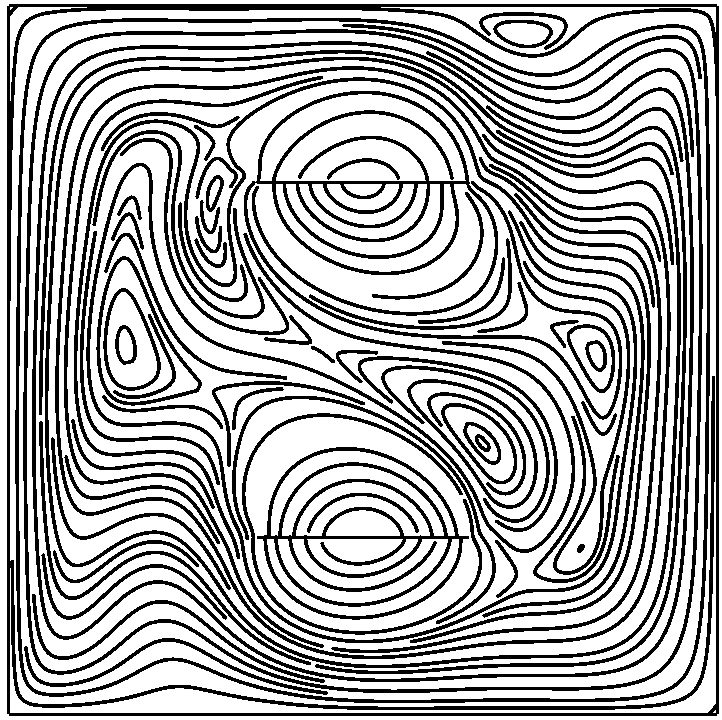}\\
				\rotatebox{90}{\hphantom{1111111111}Isotherms} & \includegraphics[width=2in,
				height=2in]{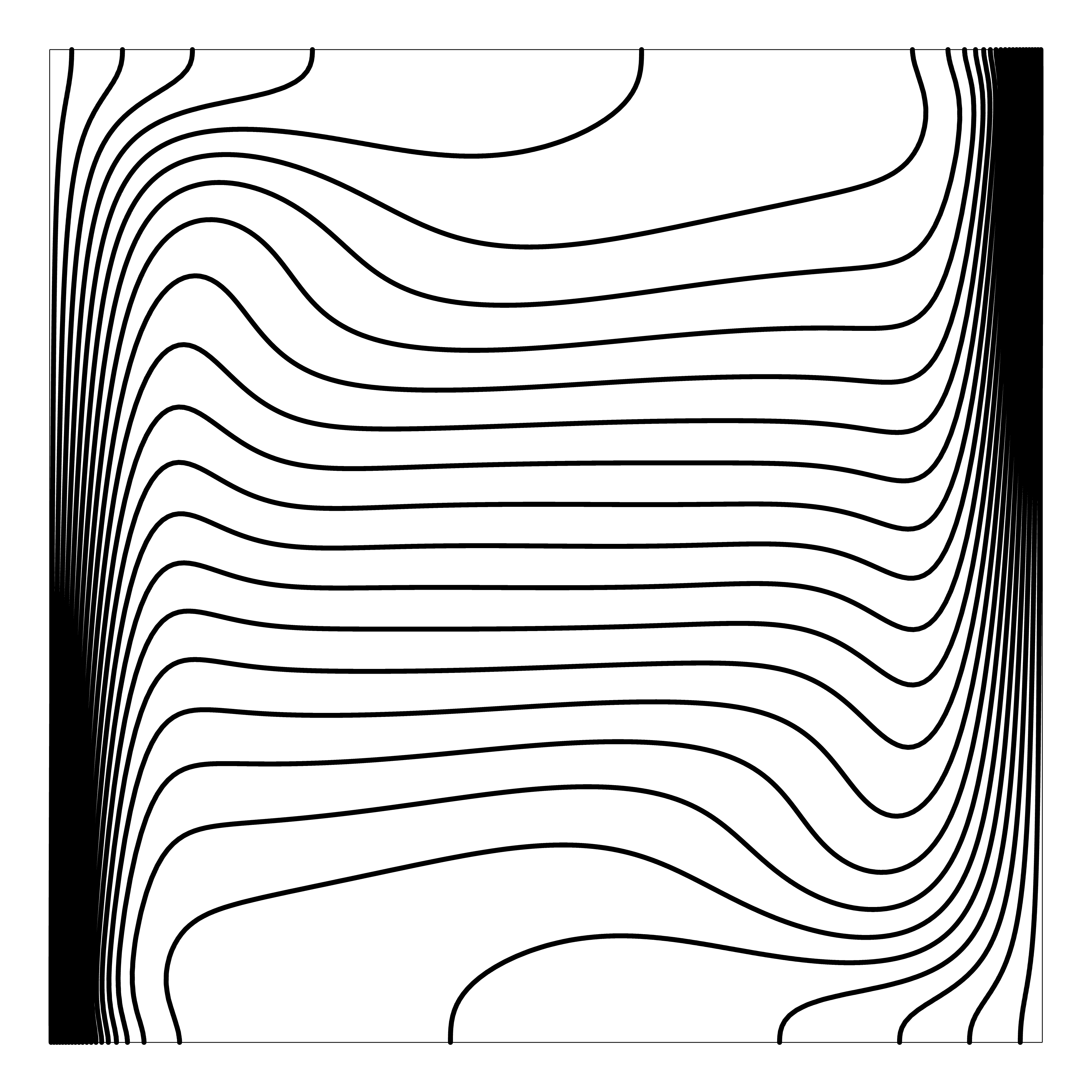} & \includegraphics[width=2in, height=2in]{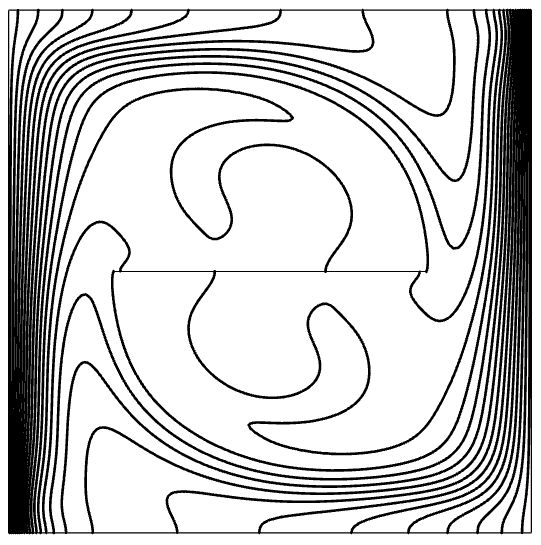} &
				\includegraphics[width=2in, height=2in]{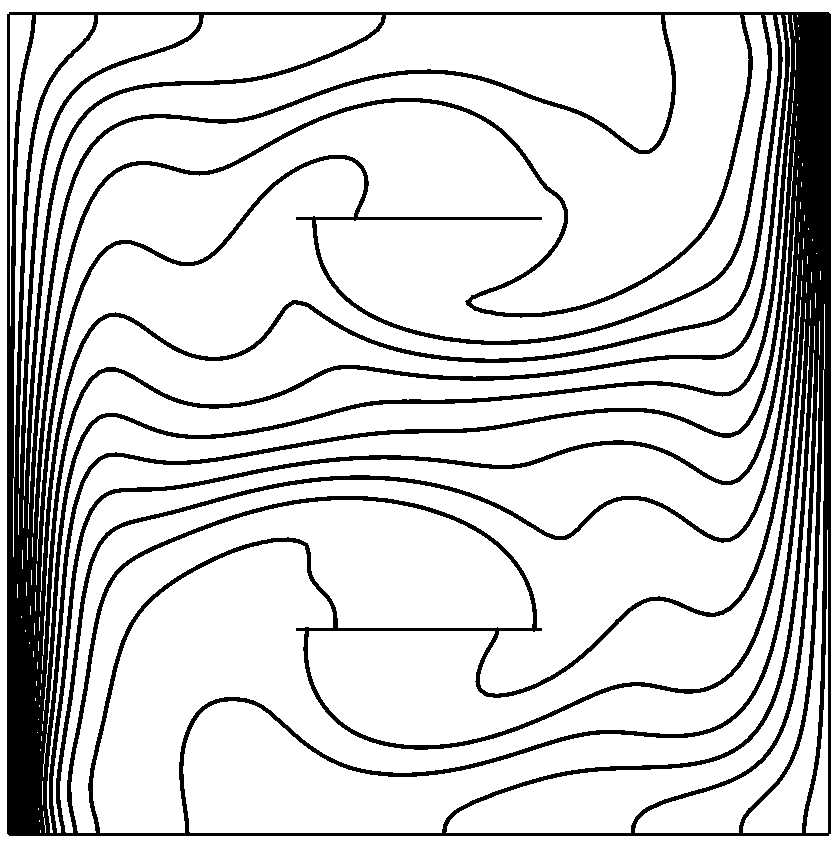}\\
				\rotatebox{90}{\hphantom{1111111111}Heatlines} & \includegraphics[width=2in,
				height=2in]{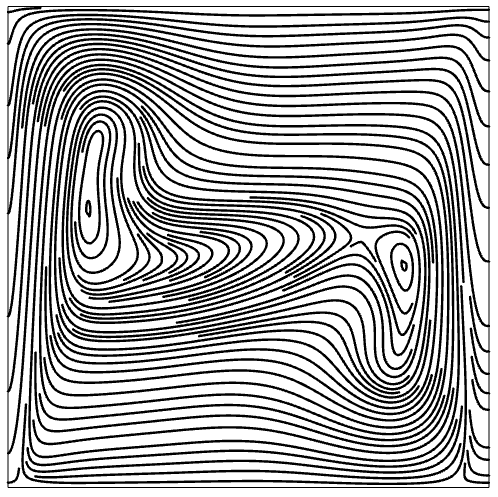} & \includegraphics[width=2in, height=2in]{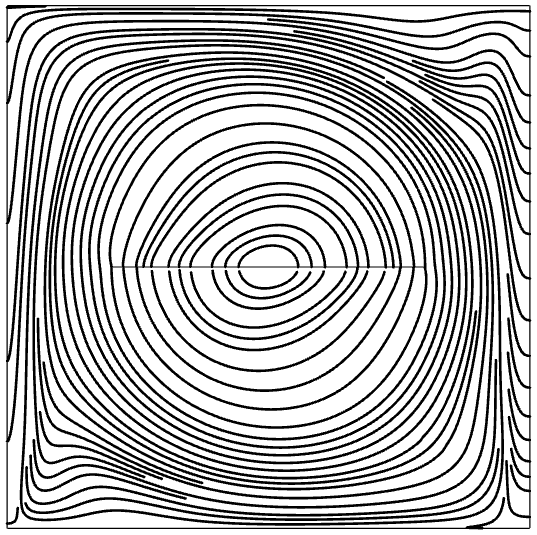} &
				\includegraphics[width=2in, height=2in]{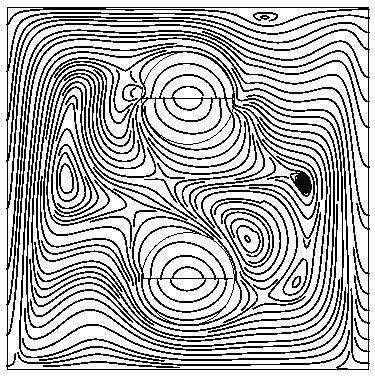}\\
			\end{tabular}
		\end{center}
		\caption{Streamlines, Isotherms and Heatlines variation for different flow modulations at $Ra=0.6 \times10^6$ at $\tau=10$}
	\end{figure}
	
	\subsection{Spectral Analysis of Spatially Average Nusselt Number ($Nu(\tau)$)}
	Rotation of the plates changes the fluid flow pattern in a periodic manner. In
	Fig. 7 FFT (\textit{Fast Fourier Transform}) has been implemented to quantify
	frequency of thermal oscillation. It has been observed that at low Rayleigh
	number ($Ra=0.0008 \times 10^6$) thermal oscillating frequency for single flow
	modulation $f_{th}$ is 2 where as for double modulation it has been found to be
	0.984. This is because due to one full rotation of single larger modulator causes
	the flow patterns and isotherms to complete two full periods. But for double
	modulator $f_{th}$ is nearly equal to the rotational frequency due to the
	opposite effects caused by the rotation of each modulator restricts the natural
	convection.
	At $Ra=0.6 \times 10^6$ for single modulation $f_{th}$ becomes 0.7115 which is
	lower than the frequency of modulator. This occurs due to instability occurring
	on the side wall boundary layer as reported by Liao \& Lin [11]. At $Ra=0.6
	\times 10^6$ for double modulation $f_{th}$ is 1 which is same as in the case for
	lower Rayleigh number.
	\begin{figure}[H]
		\begin{center}
			\begin{tabular}{c c c}
				& Single Modulation, $\frac{d}{L}=0.6$ & Double Modulation, $
				\frac{d}{L}=0.3$\\
				\rotatebox{90}{\hphantom{11111111111111}$Ra=0.0008\times10^6$} &
				\includegraphics[width=3in, height=3in]{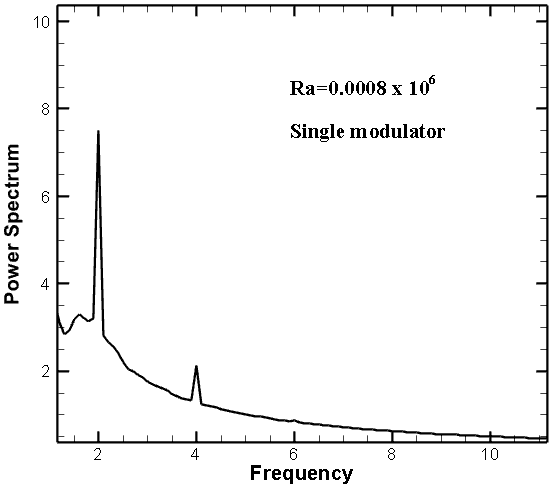} & \includegraphics[width=3in,
				height=3in]{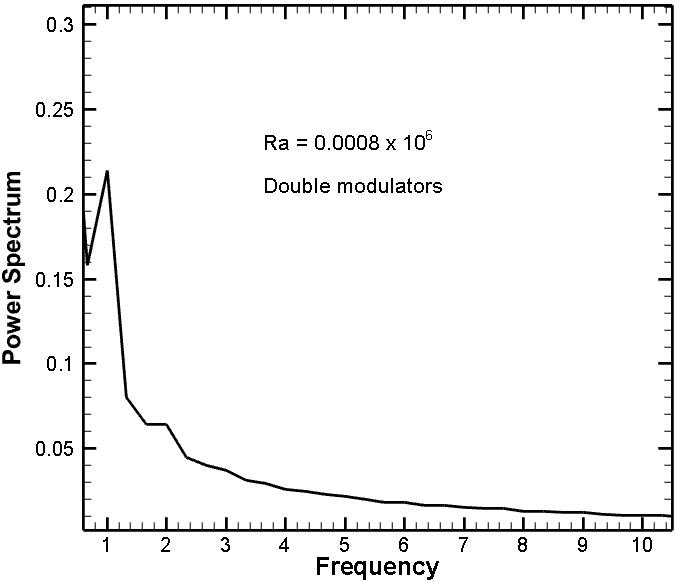}\\
				\rotatebox{90}{\hphantom{11111111111111111}$Ra=0.6\times10^6$}&
				\includegraphics[width=3in, height=3in]{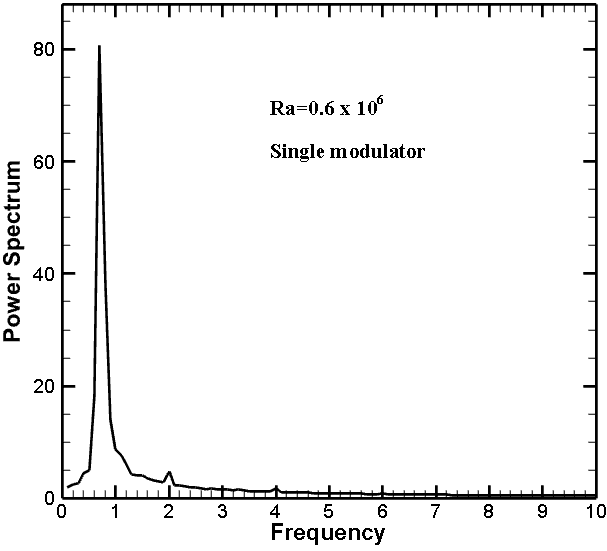} & \includegraphics[width=3in,
				height=3in]{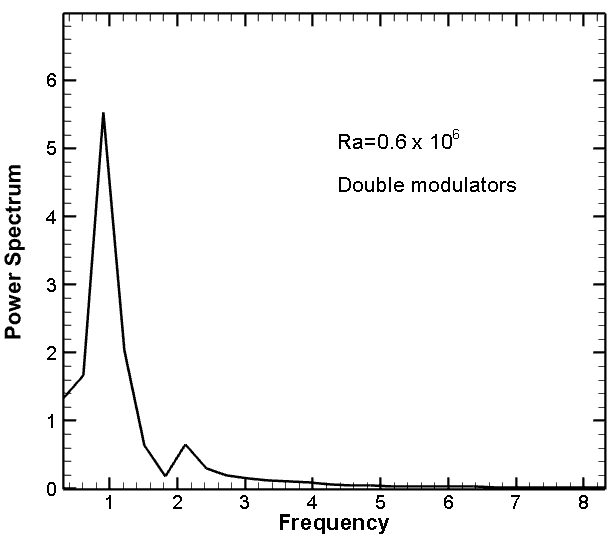}\\
			\end{tabular}
		\end{center}
		\caption{One Sided Fast Fourier Transform (FFT) Power Spectrum of Spatially
			Average Nusselt number for single/double Modulators for different Rayleigh
			numbers ($Ra$).}
	\end{figure}
	
	\section{CONCLUSIONS}
	In the present study,effect of flow modulation(via rotating flat plates) and thermal buoyancy in a differentially heated cavity
	has been investigated thoroughly using ALE finite element formulation. Results
	for different parametric conditions were graphically presented and discussed systematically.
	Based on the numerical results for different flow modulation condition such as no modulation,single modulation and double modulation(having same dynamic effect of a single modulator characterised by rotor Reynolds number) following conclusions are drawn:
	\begin{itemize}
		\item Beyond Rayeligh number $Ra_{cr}=0.41 \times 10^6$ double flow modulators
		enhance heat transfer compared to single flow modulators. At $Ra=0.6 \times 10^6$
		enhancement is 5.28\% and at $Ra=10^6$ it is 7.14\%.
		\item Thermal oscillating frequency of double flat plate is insensitive to
		Rayleigh number.
		\item At high Rayleigh numbers ($>0.41 \times 10^6$) thermal oscillation is
		prominent for single modulator.
	\end{itemize}
	
\section{REFERENCES}	
	\begin{enumerate}
		\setlength\itemsep{-0.4em}
		\item E.Lewis, J. Fluid Mech. \textbf{95}, 497–513 (1979).
		\item W.-S.Fu,C.-S.Cheng,\& W.-J.Shieh, Int.J.Heat Mass Transf. \textbf{37}, 1885–1897 (1994).
		\item N.Ghaddar,\& F.Thiele, Numer.Heat Transf. \textbf{26}, 701–717 (1994).
		\item T.Kimura,M.Takeuchi,\& K.Miyagawa, Heat Transf.-Jpn.Res. \textbf{24}, (1995).
		\item K.Khanafer \& S.Aithal, Int.Commun.Heat Mass Transf. \textbf{86}, 131–142 (2017).
		\item M.M.Billah,M.I.Khan,M.M.Rahman,M.Alam,S.Saha,and M.N.Hasan, AIP Proceedings, \textbf{1851}, 020104:1- 9 (2017).
		\item T.Kimura,M.Takeuchi,N.Nagai,Y.Kataoka,\& T.Yoshida, Heat Transfer-Asian Res.Co‐sponsored \\ Soc.Chem.Eng.Jpn.Heat Transf.Div.ASME \textbf{32}, 342–353 (2003).
		\item S.-L.Lee,J.-B.Chiou \& G.-S.Cyue, Int. J.Heat Mass Transf.\textbf{131}, 807–814 (2019).
		\item R.Glowinski,T.-W.Pan,T.I.Hesla,D.D.Joseph \& J.A Periaux, J.Comput.Phys. \textbf{169}, 363–426 (2001).
		\item H.H.Hu,N.A.Patankar \& M.Zhu, J.Comput.Phys. \textbf{169}, 427–462 (2001).
		\item C.-C.Liao \& C.-A.Lin, Phys.Fluids \textbf{27}, 013603 (2015).
	\end{enumerate}
\end{document}